\documentclass[12pt]{iopart}
\usepackage{iopams}
\usepackage{graphicx}
\input amssym.def \input amssym

\begin{document}

\title[]{Clifford groups of quantum gates, $\mbox{BN}$-pairs \\ and smooth cubic surfaces}

\author{Michel Planat$^{\dag}$ and
 Patrick Sol\'{e}$^{\ddag}$
}

\address{$^{\dag}$ Institut FEMTO-ST, CNRS, 32 Avenue de
l'Observatoire,\\ F-25044 Besan\c con, France} 

\address{$^{\ddag}$ CNRS I3S, Les Algorithmes, Euclide B,
2000 route des Lucioles,\\ BP 121, 06903 Sophia Antipolis, France} 
 

\begin{abstract}
The recent proposal (M Planat and M Kibler, Preprint 0807.3650 [quant-ph]) of representing Clifford quantum gates in terms of unitary reflections is revisited. In this essay, the geometry of a Clifford group $G$ is expressed as a $\mbox{BN}$-pair, i.e. a pair of subgroups $B$ and $N$ that generate $G$, is such that intersection $H=B\cap N$ is normal in $G$, the group $W=N/H$ is a Coxeter group and two extra axioms are satisfied by the double cosets acting on $B$. 
The $\mbox{BN}$-pair used in this decomposition relies on the {\it swap} and $\it match$ gates already introduced for classically simulating quantum circuits (R Jozsa and A Miyake, Preprint 0804.4050 [quant-ph]). The two- and three-qubit cases are related to the configuration with $27$ lines on a smooth cubic surface.

\end{abstract}

\section*{Introduction}

Euclidean real reflection groups (Coxeter groups) are an important ingredient for representing quantum computations \cite{PlanatKibler08}.
Coxeter groups are finite set of involutions and specific pairwise relations. As a result, they provide a distinguished class of quantum Boolean functions \cite{Montanaro08} possessing inherent crystallographic properties. But complex reflections are more appropriate for modeling the Clifford unitaries. For instance, the single qubit Pauli group $\mathcal{P}_1$ (generated by the ordinary Pauli spin matrices $\sigma_x$, $\sigma_y$ and $\sigma_z$) is the imprimitive reflection group $G(4,2,2)$. Its normalizer in the unitary group $U(2)$, the so-called Clifford group $\mathcal{C}_1$, is isomorphic (but is not the same as) the reflection group number $9$ in the Shephard-Todd list \cite{Kane01}. The $n$-qubit Clifford group $\mathcal{C}_n$ is the normalizer in $U(2^n)$ of the tensor product of $n$ Pauli spin matrices \cite{Clark07}. It originally appeared in the context of doubly-even self-dual classical codes \cite{Nebe01}, where it was discovered that the space of homogeneous invariants of $\mathcal{C}_n$ is spanned by the complex weight enumerators of the codes. Group $\mathcal{C}_2$ contains a maximal subgroup (of half its size) which is the Shephard-Todd group number $31$, but the connection to unitary reflection groups becomes more tenuous as far as $n\ge 3$.

In this paper, we show that Clifford groups may be seen as aggregates of Coxeter groups with the structure of $\mbox{BN}$-pairs, also named Tits systems. 
There is a compelling physical connection of the $\mbox{BN}$-pair decomposition to {\it swap} and {\it match} gates introduced in the context of classical simulations of quantum circuits \cite{JozsaMiyake08}. The $B$ group relies on the {\it swap} gates and the local component of the $n$-qubit Clifford group $\mathcal{C}_n$, while the $N$ group relies on the {\it match} gates and the topological component of $\mathcal{C}_n$. It is also noticeable that such a construction also vindicates a connection of Clifford group geometry to smooth cubic surfaces, already pointed out in our earlier work \cite{PlanatKibler08}.  

\section*{$\mbox{BN}$-pairs}

Henceforth, $G$ is finite group, $B$ and $N$ two subgroups of $G$ generating $G$, $H=B\cap N$ is a normal subgroup of $G$ and the quotient group $W=N/H$ is generated by a set $S\subset W$ of order $2$ elements (involutions). In the next section, we shall observe that such a pairing easily follows from the structure of the Clifford group $G\equiv\mathcal{C}_n$, when it is divided into its {\it local} component, the local Clifford group $B \equiv \mathcal{C}_n^{L}$, and its {\it topological} component $N \equiv\mathcal{B}_n$.

In 1962, Jacques Tits coined the concept of a $\mbox{BN}${\it -pair} for characterizing groups ressembling the general linear group over a field \cite{Tits74,Bourbaki68,Garrett97}. A group $G$ is said to have a $\mbox{BN}$-pair iff it is generated as above and two extra relations (i) and (ii) are satisfied by the double cosets \footnote{For $G$ a group, and subgroups $A$ and $B$ of $G$, each double coset is of form $A x B$: it is an equivalence class for the equivalence relation defined on $G$ by 
$$x \sim y~\mbox{if}~\mbox{there}~\mbox{are}~a\in A ~\mbox{and}~b\in B~\mbox{with}~axb=y.$$
Then G is partitioned into its $(A,B)$ double cosets.

Products of the type $sBs$ in (i) makes sense because $W$ is an equivalence class modulo $H$, and as a result is also a subset of $G$. More generally , for a subset $S$ of $W$, the product $BSB$ denotes the coset union $\bigcup_{s\in S}BsB$.  } 
  $$(\mbox{i})~~\mbox{For}~\mbox{any}~s \in S~\mbox{and}~w \in W,~ sBw \subseteq (BwB)\cup(BswB),$$
 $$(\mbox{ii})~~\mbox{For}~\mbox{any}~s \in S, sBs\nsubseteq B.$$ 

A particular example is $G=GL_n(K)$ (the general linear group over a field $K$). One takes $B$ to be the upper triangular matrices, $H$ to be the diagonal matrices and $N$ to be the matrices with exactly one non-zero element in each row and column. There are $n-1$ generators $s$, represented by the matrices obtained by swapping two adjacent rows of a diagonal matrix. More generally, any group of Lie type has the structure of a $\mbox{BN}$-pair, and $\mbox{BN}$-pairs can be used to prove that most groups of Lie type are simple. 

An important consequence of the axioms (i) and (ii) is that the group $G$ with a $\mbox{BN}$-pair may be partitioned into the double cosets as $G=BWB$. The mapping from $w$ to $C(w)=BwB$ is a bijection from $W$ to the set $B\setminus G/B$ of double cosets of $G$ along $B$ \cite{Bourbaki68}.

Let us recall that a group $W$ is a {\it Coxeter group} if it is finitely generated  by a subset $S\subset W$ of involutions and pairwise relations
$$W=\left\langle s\in S|(ss')^{m_{ss'}}=1\right\rangle,$$
where $m_{ss}=1$ and $m_{ss'}\in\left\{2,3,\ldots\right\}\cup\left\{\infty \right\}$ if $s\neq s'$. The pair $(W,S)$ is a Coxeter system, of rank $|S|$ equal to the number of generators.

The pair $(W,S)$ arising from a $\mbox{BN}$-pair is a Coxeter system. Denoting $l_s(w)$ for the smallest integer $q\ge 0$ such that $w$ is a product of $q$ elements of $S$, then (i) may be rewritten as (a) if $l_s(sw)>l_s(w)$ then $C(s w)=C(s).C(w)$, (b) if $l_s(sw)<l_s(w)$ then $C(sw)\cup C(w)=C(s).C(w)$. Such rules are the cell multiplication rules attached to the Bruhat-Tits cells $BwB$ of the Bruhat-Tits decomposition (disjoint union) $G=BWB=\bigcup_{w\in W}BwB$. 
Axiom (ii) can be rewritten as (c) for any $s \in S$, $C(s).C(s)=B \cup C(s) \neq B$.

Finally let us give the definition of a {\it split} $\mbox{BN}${\it pair}. It satisfies the two additional axioms 
$$(\mbox{iii})~~B=UH,$$ where $U$ is a normal nilpotent subgroup of $B$ such that $U\cap H =1$, and
$$(\mbox{iv})~~H=\bigcap _{n \in \mathbb N} n B n^{-1}.$$

\section*{$\mbox{BN}$-pairs from the two-qubit Clifford group}

Any action of a Pauli operator $g \in \mathcal{P}_n$ on an $n$-qubit state $\left|\psi\right\rangle$ can be stabilized by a unitary gate $U$ such that $(UgU^{\dag})U\left|\psi\right\rangle=U \left|\psi\right\rangle$, with the condition $UgU^{\dag} \in \mathcal{P}_n$. The $n$-qubit Clifford group (with matrix multiplication for group law) is defined as the normalizer of $\mathcal{P}_n$ in $U(2^n)$   

$$\mathcal{C}_n=\left\{U\in U(2^n)|U \mathcal{P}_n U^{\dag}=\mathcal{P}_n\right\}.$$

In view of the relation $U^{\dag}=U^{-1}$ in the unitary group $U(2^n)$, normal subgroups of Clifford groups are expected to play a leading role in quantum error correction \cite{PlanatKibler08,Planat08}. Let us start with the two-qubit Clifford group $\mathcal{C}_2$. The representation 
$$\mathcal{C}_2=\left\langle \mathcal{C}_1 \otimes \mathcal{C}_1, \mbox{CZ}\right\rangle$$
[where $\mbox{CZ}=\mbox{Diag}(1,1,1,-1)$ is the "controlled-$Z$" gate]
naturally picks up the local Clifford group
$$\mathcal{C}_2^{L}=\left\langle \mathcal{C}_1 \otimes \mathcal{C}_1\right\rangle=\left\langle H\otimes I, I \otimes H, P \otimes I, I\otimes P\right\rangle,$$ 

where the Hadamard gate $H:=1/\sqrt{2}\left( \begin{array}{cc}  1 & 1 \\
1& -1 \end{array} \right)$ occurs in coding theory as the matrix of the MacWilliams transform and the phase gate is $P:=\left( \begin{array}{cc}  1 & 0 \\
0& i \end{array} \right)$. The weight enumerator of Type II codes is invariant under the group of order $192$ generated by $P$ and $H,$ that is $\mathcal{C}_1$ itself \cite{McW77}. More generally the weight enumerator of genus $n$ in $2^n$ variables  is invariant under the Clifford group $\mathcal{C}_n$ \cite{Nebe01}.
The issue of efficient (classical) simulation of quantum circuits \cite{JozsaMiyake08} as well as the topological approach of quantum computation \cite{Kauf04},   suggest another decomposition of $\mathcal{C}_2$ in terms of the two-qubit gates  
$$ T:=\left(\begin{array}{cccc} 1 & 0 & 0 & 0 \\0 & 0	& 1 & 0 \\ 0 & 1 & 0 & 0 \\0 & 0 & 0 & 1\\ \end{array}\right)~\mbox{and}~R=1/\sqrt{2}\left(\begin{array}{cccc} 1 & 0 & 0 & 1 \\0 & 1	& -1 & 0 \\ 0 & 1 & 1 & 0 \\-1 & 0 & 0 & 1\\ \end{array}\right).$$

The action of gate $T$ is a {\it swap} of the two input qubits. It is straightforward to check another representation of the local Clifford group as    
$$\mathcal{C}_2^{L}=\left\langle H\otimes H, H \otimes P, T\right\rangle.$$

The action of gate $R$ is a {\it maximal entanglement} of the two input qubits. Gate $R$ is a {\it match} gate \cite{JozsaMiyake08}. It also satisfies the Yang-Baxter equation $(R \otimes I)(I \otimes R)(R \otimes I)=(I \otimes R)(R \otimes I)(I \otimes R)$ and plays a leading role in the topological approach of quantum computation \cite{Kauf04}. It was used in our earlier work to define the {\it Bell group}
$$\mathcal{B}_2=\left\langle H\otimes H, H \otimes P, R\right\rangle.$$

Both groups $\mathcal{C}_2^{L}$ and $\mathcal{B}_2$ are subgroups of order $4608$ (with index $20$) and $15360$ (with index $6$) of the Clifford group. The latter may be represented as $\mathcal{C}_2=\left\langle H\otimes H, H \otimes P, \mbox{CZ}\right\rangle.$ 

\subsection*{The search of the $\mbox{BN}$-pairs}

Clearly, the Clifford group is generated by the local Clifford group $\mathcal{C}_2^{L}$ and Bell group $\mathcal{B}_2$. Their intersection is the Pauli group $\mathcal{P}_2$, of order $64$, that is isomorphic to the central product $E_{32}^+*\mathbb{Z}_4$ (where $E_{32}^+$ is the extraspecial $2-$group of order $32$ and type $+$). The Pauli group $\mathcal{P}_2$ is normal in the Clifford and Bell groups but neither of the quotient groups $\mathcal{C}_2^{L}/\mathcal{P}_2$ and $\mathcal{B}_2/\mathcal{P}_2 \cong \mathbb{Z}_2 \times S_5$ is a Coxeter group, so that the pair $(\mathcal{C}_2^{L},\mathcal{B}_2)$ cannot be of the $\mbox{BN}$-type.

Let us search a $\mbox{BN}$-pair candidate by selecting the subgroup $N \equiv \mathcal{B}_2$ and reducing the size of $\mathcal{C}_2^{L}$ to a subgroup $B$ so that the intersection group $H=N\cap B$ is a subgroup of $B$ and $N/H$ is a Coxeter group. One gets 
$$B\cong W(F_4),~N\equiv \mathcal{B}_2,~H \equiv Z(\mathcal{B}_2)\cong \mathbb{Z}_8~\mbox{and}~W\cong W(D_5),$$
in which $B$ is the unique subgroup of $\mathcal{C}_2^{L}$ which is both of order $1152$ and isomorphic to the Coxeter group $W(F_4)$ of type $F_4$ (the symmetry group of the $24$-cell), $N$ is $\mathcal{B}_2$, $H$ is the center $Z(\mathcal{B}_2)$ and $W$, of order $1920$, is isomorphic to the Coxeter group $ W(D_5)$ of type $D_5$.

The above pair of groups is of the  $\mbox{BN}$ type seeing that conditions (i) and (ii) are satisfied. Axiom (i) directly follows from the Coxeter group structure of $W$. For (ii), which is equivalent to (c), it is enough to discover an element in the double coset $C(s)$ which does not lie in group $B$. Elements of the coset $C(s)=BsB$ arise from elements of the coset $\mathcal{C}_2^{L} g  \mathcal{C}_2^{L}$, $g \in \mathcal{B}_2$. The latter coset contains the entangling match gate $R'=T R  T$, which lies in $\mathcal{B}_2$ but not in $\mathcal{C}_2^{L}$.  Thus (c) is satisfied.
 The $\mbox{BN}$ pair does not split because there is no normal subgroup of order $|B|/|H|=144$ within the group $B$.
 
\subsection*{A split $\mbox{BN}$-pair}

A further structure may be displayed in the two-qubit Clifford group. Let us denote $\hat{G}$ the central quotient of the derived subgroup of $G$. One immediately checks that $\hat{\mathcal{C}_2}=\langle \hat{\mathcal{C}}_2^{(L)},\hat{B_2}\rangle \cong U_6$, $\hat{B_2}\cong M_{20}$ and $\hat{\mathcal{C}}_2^{(L)} \cong \hat{W}(F_4)$. Group $U_6=\mathbb{Z}_2^4 \rtimes A_6$, of order $5760$, appears in several disguises. The full automorphism group of the Pauli group $\mathcal{P}_2$ possesses a derived subgroup isomorphic to $U_6$ (see relation (7) in \cite{PlanatKibler08}). Geometrically, it corresponds to the stabilizer of an hexad in the Mathieu group $M_{22}$ (see Sec 4.2 in \cite{PlanatKibler08}). Group $M_{20}=\mathbb{Z}_2^4 \rtimes A_5$, of order $960,$ is isomorphic to the derived subgroup of the imprimitive reflection group $G(2,2,5)$ (see Sec 3.5 in \cite{PlanatKibler08}). Incidentally, $M_{20}$ is the smallest perfect group for which the set of commutators departs from the commutator subgroup \cite{Planat08}. Remarkably, the group $\hat{\mathcal{C}_2}$ forms the split $\mbox{BN}$-pair
$$B\equiv \hat{\mathcal{C}}_2^{(L)},~N\equiv \hat{\mathcal{B}_2},~H\equiv \tilde{\mathcal{P}}_2 \cong \mathbb{Z}_2^4,~W\cong A_5,~\mbox{and}~U\cong\mathbb{Z}_3^2.$$     

\section*{$\mbox{BN}$-pairs from the three-qubit Clifford group}

The local Clifford group
$$\mathcal{C}_3^{(L)}=\left\{\mathcal{C}_1 \otimes \mathcal{C}_1 \otimes \mathcal{C}_1 \right\},$$
and the three-qubit Bell group 
$$\mathcal{B}_3=\left\langle H\otimes H \otimes P, H \otimes R, R \otimes H\right\rangle,$$
are subgroups of index $6720$ and $56$, respectively, of the three-qubit Clifford group (of order $743~178~240$). It may be generated as
$$\mathcal{C}_3=\left\langle H\otimes H \otimes P, H \otimes \mbox{CZ}, \mbox{CZ} \otimes H\right\rangle.$$

The central quotients $\tilde{\mathcal{C}_3}$ and $\tilde{\mathcal{B}_3}$ may be expressed  as semi-direct products
$$\tilde{\mathcal{C}_3}=\mathbb{Z}_2^6 \rtimes W'(E_7)~~\mbox{and}~~\tilde{\mathcal{B}_3}=\mathbb{Z}_2^6 \rtimes W'(E_6), $$
in which $W'(E_7)\equiv \mbox{Sp}(6,2)$ and $W'(E_6)$ are the reflection groups of type $E_7$ and $E_6$, respectively \cite{PlanatKibler08}.
Following the intuition gained from the previous section, one immediately gets the {\it non-split} \footnote{The pair is not split since the quotient group $V$ is not normal in $\tilde{\mathcal{C}}^{(L)}_3$, not nilpotent and $V \cap H \neq 1$.    } \mbox{BN}-pair 
$$B\equiv \tilde{\mathcal{C}}^{(L)}_3,~N\equiv \tilde{\mathcal{B}_3},~H\equiv \tilde{\mathcal{P}}_3 \cong \mathbb{Z}_2^6, ~W\cong W'(E_6),~\mbox{and}~\tilde{\mathcal{C}}^{(L)}_3/H \equiv V,$$ 
in which $V\cong S_3^3$ ($S_3$ is the symmetric group on three letters).     

\subsection*{\mbox{BN}-pairs and a smooth cubic surface}

The occurence of reflection groups $W(F_4)$ and $W(D_5)$ in the decomposition of the two-qubit Clifford group, and of $W(E_6)$ in the decomposition of the three-qubit Clifford group, can be grasped in a different perspective from the structure of a {\it smooth cubic surface} $\mathcal{S}$ embedded into the three-dimensional complex projective space $\mathbb{P}^3(\mathbb{C})$ \cite{Hunt95}. The surface contains a maximum of $27$ lines in general position and $45$ sets of tritangent planes. The group of permutations of the $27$ lines is $W(E_6)$, the stabilizer of a line is $W(D_5)$ (observe that $|W(E_6)|/|W(D_5)|=27$) and the stabilizer of a tritangent plane is $W(F_4)$. Thus the $\mbox{BN}$-pairs happens to be reflected into the geometry of such a cubic surface. 

Other \lq\lq coincidences " occur as follows. The number $216$ of pairs of skew lines of $\mathcal{S}$ equals the cardinality of the quotient group $V$ entering in the decomposition of $\tilde{\mathcal{C}}_3$. There are $36$ double sixes, each one stabilized by the group $g_6:=A_6.\mathbb{Z}_2^2$ of order $1440$ (the symbol $.$ means that the group extension does not split). The latter group can be displayed in the context of the two-qubit Clifford group.
Let us observe that the quotients of $\mathcal{C}_2$ and $\mathcal{B}_2$ by the Pauli group $\mathcal{P}_2$ are isomorphic to $g_6$ and $g_5:=A_5.\mathbb{Z}_2^2$, respectively. For three-qubits, one checks that the quotients of $\mathcal{C}_3$ and $\mathcal{B}_3$ by the Pauli group $\mathcal{P}_3$ are isomorphic to $W(E_7)$ and $W(E_6)$. Groups $W'(E_6)$, $W(D_5)$, $W(F_4)$ and $g_6$, which correspond to the permutations of the $27$ lines, the stabilizer of a line, a tritangent plane and a double six, respectively, are among the six maximal subgroups of $W(E_6)$. The remaining two are of order $1296$ and index $40$, corresponding to the size of double cosets $BwB$, $B\cong W(F_4)$ and $w \in W \cong W(D_5)$, in the $\mbox{BN}$-pair decomposition of the two-qubit Clifford group.  

To conclude, a smooth cubic surface is a particular instance of a $K_3$ surface, a concept playing a founding role in string theory. Further work is necessary to explore the interface between quantum computing, graded rings and $K_3$ surfaces \cite{Altinok02}.  

\section*{Acknowledgements}
The first author warmly acknowledges Richard Jozsa and Noah Linden for the impetus given to this work, following his recent visit at the Department of Computer Science, Bristol. He also received a constructive feedback by Maurice Kibler and Metod Saniga.

\end{document}